\documentclass[a4paper]{llncs}
\usepackage{amsmath,amssymb,amsfonts,cancel}
\usepackage{graphicx}
\usepackage{algorithm} 
\usepackage{algpseudocode}
\usepackage[font={small}]{caption}
\usepackage{sidecap}
\usepackage{url}

\begin{document}
\title{Semi-supervised Classification for Dynamic Android Malware Detection}

\author{Li Chen\thanks{First and corresponding author's email address is li.chen@intel.com. Li Chen is Data Scientist in Privacy and Security Lab in Intel Labs, Hillsboro, OR 97124.},  Mingwei Zhang\thanks{Mingwei Zhang is Research Scientist in Privacy and Security Lab in Intel Labs, Hillsboro, OR 97124.}, Chih-Yuan Yang\thanks{Chih-Yuan Yang is Research Scientist in Privacy and Security Lab in Intel Labs, Hillsboro, OR 97124.}, Ravi Sahita\thanks{Ravi Sahita is Principal Engineer in Privacy and Security Lab in Intel Labs, Hillsboro, OR 97124.}}
\institute{Security and Privacy Research, Intel Labs}
\maketitle

\begin{abstract}

A growing number of threats to Android phones creates challenges for malware detection. Manually labeling the samples into benign or different malicious families requires tremendous human efforts, while it is comparably easy and cheap to obtain a large amount of unlabeled APKs from various sources. Moreover, the fast-paced evolution of Android malware continuously generates derivative malware families. These families often contain new signatures, which can escape detection when using static analysis. These practical challenges can also cause traditional supervised machine learning algorithms to degrade in performance.

In this paper, we propose a framework that uses model-based semi-supervised (MBSS) classification scheme on the dynamic Android API call logs. The semi-supervised approach efficiently uses the labeled and unlabeled APKs to estimate a finite mixture model of Gaussian distributions via conditional expectation-maximization and efficiently detects malwares during out-of-sample testing. We compare MBSS with the popular malware detection classifiers such as support vector machine (SVM), $k$-nearest neighbor (kNN) and linear discriminant analysis (LDA). Under the ideal classification setting, MBSS has competitive performance with 98\% accuracy and very low false positive rate for in-sample classification. For out-of-sample testing, the out-of-sample test data exhibit similar behavior of retrieving phone information and sending to the network, compared with in-sample training set. When this similarity is strong, MBSS and SVM with linear kernel maintain 90\% detection rate while $k$NN and LDA suffer great performance degradation. When this similarity is slightly weaker, all classifiers degrade in performance, but MBSS still performs significantly better than other classifiers.

\end{abstract}


\section{Introduction}


%

Android had dominated the mobile operating system and accounted for over 86\% of global shipment and operating system market share in 2016 \cite{gartner} \cite{idc}. The Android’s official application store, Google Play, provides over 2.7 million applications \cite{AppBrain} and total download count had reached 65 billion times in one year period \cite{Google}. The popularity of Android operating system has made it a lucrative target for cybercriminals.
According to McAfee’s 2016 mobile threat report, there have been 9 millions of malware found from app stores cross 190 countries globally in three month period. And 3 million devices were affected over 6 month’s period \cite{mc}.
The average monthly infection rate among smartphones increased to 0.49 percent in the first half of 2016. This is a 98 percent surge from the 0.25 percent in the second half of 2015 based on Nokia Threat Intelligence lab’s report \cite{nokia}.   


The rampant evolution of Android malware makes it more sophisticated to avoid detection and more difficult to classify using the commonly-used machine learning algorithms. Human effort of labeling the malware as malicious or benign cannot keep up with the pace of the voluminous generation of Android malware, resulting in an imbalance of much more unlabeled data than labeled data. For supervised learning algorithms, this causes potential difficulties, since these algorithms are constructed merely on the labeled dataset or training data, but are desired to have reasonably good performance on the greater amount of unlabeled data. 


Furthermore, due to this imbalance of unlabeled and labeled data, the distribution observed from the labeled data can be different from the actual data distribution. This is seen in malware detection, as study suggests that malware family exhibit polymorphic behaviors. Translated into machine learning, it implies that at testing phase, the test data can be similarly distributed as the training data, but is not identically distributed as the training data. Traditional supervised machine learning algorithms can suffer performance degradation when tested on samples that do not distribute identically as the training data. Therefore, in order to achieve a robust malware detection rate for out-of-sample malware detection, it is desired to have a machine learning based malware detector that takes advantage of both unlabeled and labeled data, and maintains a steady performance for out-of-sample test analysis.

To address the above challenges, in this paper, we propose a framework of utilizing model-based semi-supervised (MBSS) classification on the dynamic behavior data for Android malware detection. We focus on detecting malicious behavior at runtime by using dynamic behavior data for our analysis. The main advantage of semi-supervised classification is the strong robustness in performance for out-of-sample testing. The model-based semi-supervised classification uses both labeled and unlabeled data to estimate the parameters, since unlabeled data usually carries valuable information on the model fitting procedure. 

Specifically, we use mixture modeling to achieve the classification task. Mixture modeling is a type of machine learning technique, which assumes that every component of the mixture represent a given set of observations in the entire collection of data. Gaussian mixture modeling is the most popularly used applied and studied technique. Model-based mixture modeling uses a mixture of Gaussian distributions to develop clustering, classification and semi-supervised learning methods \cite{banfield1993model}, \cite{fraley2002model}, \cite{fraley1998many}, \cite{dean2006using}. 

We run the Android applications in our emulator infrastructure and harvest the API calls at runtime. Our framework efficiently uses the labeled and unlabeled behavior data to estimate a set of finite mixture models of Gaussian distributions via conditional expectation-maximization, and uses the Bayesian information criterion for model selection. We compare MBSS with the popular malware detection classifiers such as support vector machine (SVM), $k$-nearest neighbor (kNN) and linear discriminant analysis (LDA). We demonstrate that MBSS has competitive performance for in-sample classification, and maintains strong robustness when applied for out-of-sample testing. We consider semi-supervised learning on dynamic Android behavior data a practical and valuable addition to Android malware detection.

The rest of the paper is structured as follows. In Section \ref{sec:related_work}, we describe related work of machine learning used in Android malware detection. In Section \ref{sec:android_malware_emulator_imp}, we describe our infrastructure and implementation for our Android malware emulator, which collects the dynamic behavioral data for our analysis. In Section \ref{sec:mbss}, we describe in detail the model-based semi-supervised classifier. In Section \ref{sec:result}, we present the results of our proposed methodology for three sets of experiments to demonstrate its effectiveness for out-of-sample testing. We conclude our paper with summary and discussion in Section \ref{sec:conclusion}.


%
\section{Related Work}\label{sec:related_work}
The intersection of artificial intelligence and statistics provides machine learning the foundation of probabilistic models and data-driven parameter estimation. Machine learning tasks can be categorized into supervised learning, unsupervised learning and semi-supervised learning. In supervised learning, the learning task is to classify the labels of incoming data samples given the observed training data and their labels. Traditional methods include nearest neighbor \cite{cover1967nearest}, support vector machine \cite{cortes1995support}, decision tree \cite{safavian1991survey}, sparse representation classifier \cite{wright2009robust} \cite{chen2016robust}, etc. In unsupervised learning, the learning problem is to group the observations into categories based on a chosen similarity measure. Typical methods include K-means, expectation-maximization \cite{dempster1977maximum}, \cite{moon1996expectation}, model-based clustering \cite{fraley2002model}, \cite{chen2014stochastic}, hierarchical clustering \cite{johnson1967hierarchical}, spectral clustering \cite{ng2001spectral} \cite{lyzinski2015spectral}, and so on. 
Semi-supervised is between supervised and unsupervised learning, where the learning problem makes use of the unlabeled for training and updates the model using both the labeled and unlabeled data. Typical methods include Gaussian mixture models \cite{mclachlan2004finite}, hidden Markov model, low-density separation \cite{chapelle2005semi}, and so on. 

There are various techniques used to detect malware on Android system. They generally could be classified as static and dynamic analysis. Static analysis techniques focus on the information from Android application package (APK) such as manifest, resources, code binary, etc., while dynamic analysis techniques focus on dynamic behaviors collected during APK execution.

Previous work such as CHEX \cite{lu2012chex} statically checks the code logic to prevent component hijacking attacks that might cause privacy leakage. MAMADROID \cite{mariconti2016mamadroid} and DroidAPIMiner \cite{aafer2013droidapiminer} both use Android API invocations in static code to detect malware. While static techniques are more prevalent among anti-malware companies, they are known to fail to detect malwares with sophisticated evasion techniques like dynamic code loading or code obfuscation. MYSTIQUE-S \cite{xue2017auditing} is a sample malware, which can select attack features at runtime, and download malicious payloads dynamically. It bypasses static detection and can only be detected by dynamic monitoring tools such as Droidbox \cite{droidbox}. 
 

DroidHunter \cite{shabtai2010google}, FLEXDROID \cite{seo2016flexdroid} and Going native \cite{afonso2016going} are a few examples on dynamic detection techniques. DroidHunter is able to detect malicious code loading by hooking critical android runtime API. FLEXDROID proposes an extra component in Android app manifest, which allows developers to set permissions on native code. Going Native generates policies for native code and limits malicious behaviors. On the other hand, several dynamic monitoring tools are proposed on Android malware detection. 
Droidbox \cite{droidbox} and Droidmon \cite{droidmon} are both based on instrumentations in android runtime. 
Droidbox relies on source code modification of Android Open Source Project (AOSP), thus suffers from huge engineering works of porting various Android versions. Droidmon instead is based on Xposed Framework, which is a framework for modules that can change the behavior of APK without modifying it 

 
One downside of dynamic analysis is the limited code coverage as some behaviors only exposed under specific conditions such as user interactions or sensor data. IntelliDroid \cite{wong2016intellidroid} tries to solve this problem using symbolic execution. It leverages a solver to generate targeted input to trigger malicious code logic and increase the coverage. However, its implementation is bound to Android version, and may suffer from the porting issue.
 

Similar to Droidmon, our instrumentation framework Android Emulator (AE) is based on Xposed Framework for dynamic API monitoring. Also several components, such as the UI automation tool and sensor emulators, were incorporated in AE to enhanced coverage of dynamic behavior. Tracing logs of APK execution were harvested for future machine learning analysis.

\section{ Data Generation and Preparation}\label{sec:android_malware_emulator_imp}
\begin{figure}
\center
\includegraphics[width = 8.8cm, height=5cm]{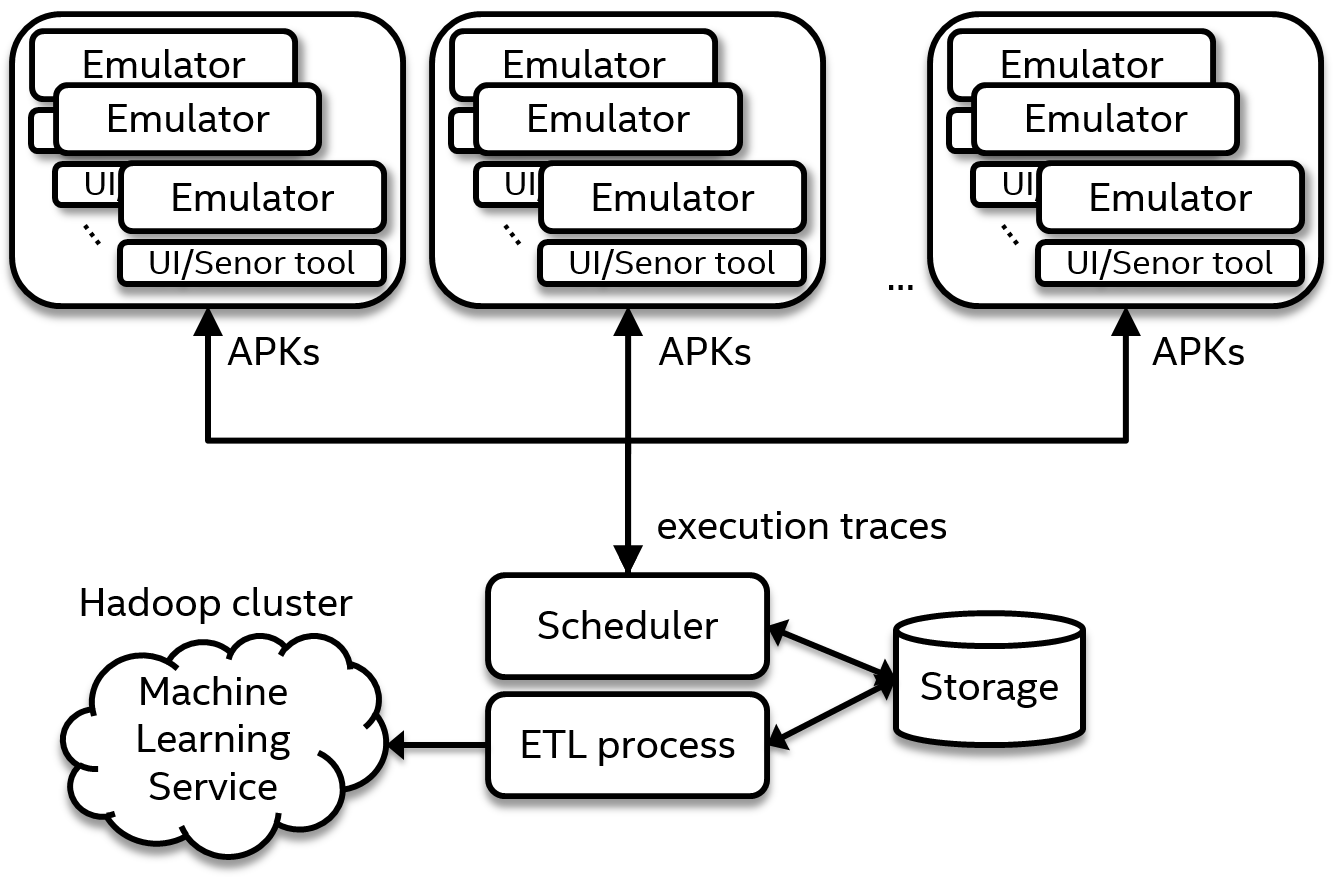}
\caption{Dynamic Instrumentation System Architecture.}
\label{fig:emulator}
\end{figure}

Malicious and benign APKs were downloaded from VirusTotal and Google Play respectively. Both samples were uploaded to our emulator infrastructure for execution, as seen in Figure \ref{fig:emulator}. The Android emulator (AE) runs in a emulator machine. Each machine may contains multiple emulators. The downloaded APKs were dispatched by the scheduler and installed to new emulator instances for execution. Since most of Android applications are UI-based, merely launching the application may be insufficient to expose its behaviors. The automation tool is developed to provide simulated human interactions, such as clicks, and sensor events, such as GPS location. This tool could navigate the UI automatically without human intervention. We harvest applications behaviors through Android Debug Bridge (ADB) \cite{adb} and aggregate them into the disk. 


To address the trade-off between efficiency and completeness, we set our experiment time for each APK as 10 minutes. To harvest the API calls at runtime, our customized emulator has pre-installed Xposed framework \cite{exposed} which could potentially hook any API invocations to android runtime. Our Xposed component is running along with each application instance and intercept and print each API invocation through ADB \cite{adb}, which is further harvested by a dedicated out-of-box process. Since Android application is mostly UI-based, launching the application is insufficient to harvest enough behavior. To cope with it, we used a UI automation tool playing as a robot to dynamically click the running APK inside emulator. As a result, we could navigate most of the application logic with largely reduced the human effort.

We capture dynamic behavior of each android application by executing it in our emulator. The dynamic data consists of a time sequence of API calls made by an application to the Android runtime. Since current Android runtime export more than 50K APIs, for efficiency, we carefully select 160 API calls that are critical to change Android system state such as sending short messages, accessing website, reading contact information, etc. Our selection of Android API function comes from the union of function set selected by three well-known open source projects: AndroidEagleEye \cite{eagleeye}, Droidmon \cite{droidmon}, and Droidbox \cite{droidbox}. We believe our API selection set is sufficient to cover all critical Android malicious behaviors at runtime. For each API call, we capture the API class name and its function name.

As a result, our data consists of the dynamic traces of the APKs as samples, and the feature space of our data set is the collection of APIs, where in each feature, we denote the existence of the API in the sample APK. That is, suppose the unique number of API calls is $d$, then a sample APK is represented by $x = (0,0,..,1,..,1,..,0)\in \{0,1\}^d$, where 1 denotes the existence of an API call and 0 otherwise. 

\section{Model-based semi-supervised classification} \label{sec:mbss}
Even if a classifier achieves high classification accuracy and low false positive rate for in-sample testing, its good performance may not be extended to out-of-sample testing. Semi-supervised learning classification uses both the labeled and unlabeled to update the classifier, taking advantage of the complete dataset and thus achieving more accurate classification.
Here we propose to use the model-based semi-supervised (MBSS) classification algorithm \cite{dean2006using}, \cite{russell2014up}, \cite{fraley2002model}. We start with formulating the semi-supervised classification task, then explain the conditional expectation-maximization algorithm used for solving the maximum likelihood problem, and describe the model selection criterion. 

\subsection{The Semi-supervised Learning Task}

In the supervised setting, we define $[K] = \{1,2,...,K\}$ for a given positive integer $K$. Suppose $(X,Y) \sim F_{XY}$, where $X$ is a feature vector in $\mathbb{R}^d$ and $Y \in [K]$ is the class label, and $F_{XY}$ is the joint distribution of $X$ and $Y$. Further suppose we observe the independently identically distributed training data $T_n=(\{\mathcal{X}_n\}, \{\mathcal{Y}_n)\} := \{(X_1, Y_1), (X_2, Y_2), ...,(X_n, Y_n)\} \overset{i.i.d}{\sim} F_{XY}$. The task is to classify the class membership of sample $X$ via defining a classifier $g: X \rightarrow Y$ \cite{devroye2013probabilistic}. 

In the unsupervised setting, we observe independently identically distributed feature vectors $X_1, X_2, ...,X_n$ where each $X_i$ is a random variable for some probability space. The task is to cluster the samples into groups based a chosen metric of similarity. 

The semi-supervised setting is between supervised and unsupervised learning. Denote the incoming data by $\mathcal{X}_m = \{X_{n+1}, ..., X_{n+m} \}$ with unknown labels and use the same notation $T_n$ as the training data. The task is to learn a classifier $g: \mathcal{X} \rightarrow \mathcal{Y}$ such that this classifier predicts the labels of $\mathcal{X}_m$ via learning from the complete data rather than the labeled training data alone. 

\subsection{Model-based Semi-supervised Learning Framework}
In model-based approach, the data $x$ is assumed to be distributed from a mixture density $f(x)_ = \sum_{k=1}^K\pi_k f_k(x)$, where $f_k(\cdot)$ is the density from the $k$-th group and $\pi_k$ is the probability that an observation belongs to group $k$. Each component is Gaussian, which is characterized by its mean $\mu_k$ and covariance matrix $\Sigma_k$. The probability density function for the $k$-th component is thus 
\begin{equation}
f_k = f_k(x, \mu_k, \sigma_k) = \frac{\exp(-\frac{1}{2}(x-\mu_k)^T\Sigma_k^{-1}(x-\mu_{k}))}{\sqrt{\det(2\pi\Sigma_k)}}.
\end{equation} 


The mean $\mu = (\mu_1, ..., \mu_K)$, covariance matrix $\Sigma = (\Sigma_1, ..., \Sigma_K)$ and the population distribution $\pi = (\pi_1, ..., \pi_K)$ are the parameters to be estimated from the mixture models. Here we use maximum likelihood estimation, a method of finding the parameters to maximize the probability of obtaining the observations given the parameters.  

Denote $\theta := (\mu, \Sigma)$ as all the parameters in the Gaussian components. Hence we want to estimate $\pi$ and $\theta$. Denote $\mathcal{X}_n = \{X_1, ..., X_n\}$ as the training data and $\mathcal{Y}_n = \{Y_1, ..., Y_n\}$ as the  training labels, where for the i-th observation, denote $Y_{ik} = 1$ if the observation comes from group $k$ and 0 otherwise. Denote the unknown labels of the unlabeled data as $\mathcal{Y}_M = \{Y_{n+1},...,Y_{n+m}\}$.
The likelihood of the training data is 
\begin{equation}
L_T(\pi, \theta|\mathcal{X}_n, \mathcal{Y}_n, \mathcal{X}_m) = \Pi_{i=1}^n\Pi_{k=1}^K[\pi_kf(X_i|\theta_k)]^{Y_{i,k}}\pi_{j=n+1}^{n+m}\sum_{k=1}^K\pi_kf(X_j|\theta_k).
\end{equation} 
To estimates the unknown parameters $\pi$ and $\theta$, one calculates log-likelihood as $l(\pi, \theta) = \log L_T(\pi, \theta)$ and uses the EM algorithm \cite{dempster1977maximum}, \cite{moon1996expectation} to maximize the log-likelihood. 

In our framework, we apply model-based semi-supervised classification using both labeled and unlabeled Android behavioral data to develop the classification decision for the unlabeled data. The likelihood of the complete-data consisting of labeled and unlabeled data is 
\begin{equation}
L_C(\pi, \theta| \mathcal{X}_n, \mathcal{Y}_n, \mathcal{X}_m, \mathcal{Y}_m) = \Pi_{i=1}^n\Pi_{k=1}^K[\pi_kf(x_i|\theta_k)^{Y_{ik}}]\Pi_{j=n+1}^{n+m}\Pi_{k=1}^K[\pi_kf(X_j|\theta_k)]^{Y_{jk}}.
\end{equation}
Essentially we treat the data with unknown labels as missing data to include them in the complete likelihood.
To estimate the unknown parameters and maximize the log-likelihood of the complete data, we use the conditional expectation-maximization (CEM) algorithm \cite{jebara1998maximum}, which is similar to the expectation-maximization (EM) algorithm. 

\subsection{The Conditional Expectation-Maximization Algorithm}
The conditional expectation-maximization (CEM) \cite{jebara1998maximum} algorithm is used to solve the likelihood maximization on the complete data. We note that CEM is similar to EM, except we update the classification using both labeled and unlabeled data. We denote the parameter estimates at the $g$-th iteration by $\hat{\pi}^g, \hat{\theta}^g = (\hat{\mu}^g, \hat{\Sigma}^g)$ and the estimated label on the test data at the $g$-th iteration by $\hat{Y}^g_m$. 
\begin{itemize}
\item Step 1: \textbf{Initialization}. Set $g = 0$ and initiate starting estimates $\hat{\pi}^0, \hat{\theta}^0$ using model-based discriminant analysis estimates from the parameters of the model, as described in \cite{fraley2002model}.
\item Step 2: \textbf{Expectation}. While chosen stopping criterion is not satisfied, increase iteration $g$. Calculate the expected value of the unknown labels via
\begin{equation}
w_{jk} = \frac{\pi_kf(X_j|\hat{\theta}^g_k)}{\Sigma_{k=1}^K \hat{\pi}^g_kf(X_m|\hat{\theta}^g_k)},
\end{equation}
where $k \in [K]$ denotes the index of the number of class memberships and $j = n+1, ..., n+m$ denotes the index on the unlabeled data.

\item Step 3: \textbf{Maximization}. The estimated labels after the $g$-th iteration is $\hat{Y}^{g+1}_{jk}:= \text{sign}(w_{jk} - w_{jk'})$ for all $k\neq k'$.
Estimate the mixture parameters by
\begin{equation}
\hat{\pi}_k^{g+1} = \frac{\sum_{i=1}^nl_{ik} + \sum_{j=n+1}^{n+m}\hat{Y}^{g+1}_{jk}}{n+m},
\end{equation}
\begin{equation}
\hat{\mu}_{k}^{g+1} = \frac{\sum_{i=1}^nl_{ik}X_i + \sum_{j=n+1}^{n+m}\hat{Y}_{jk}^{g+1}}{\sum_{i=1}^{n}l_{ik}+\sum_{j=n+1}^{n+m}\hat{Y}_{jk}^{(g+1)}}.
\end{equation}
To estimate the covariance, we apply eigenvalue decomposition on the covariance matrix $\Sigma_k$, i.e., $\Sigma_k = \lambda_kD_kA_kD^T_k$. Depending on which covariance structure is, seen in Section 4.4, different constraints are imposed on the covariance matrix. \cite{bensmail1996regularized}, \cite{celeux1995gaussian} describe in detail how each covariance matrix is estimated according to the structure.
\item Step 4: \textbf{Convergence}. Stop until stopping criteria is met.
\end{itemize}

In this study, the stopping criterion is set when the current value of the log-likelihood is within $10^{-5}$ close to the estimated final converged value. Upon convergence, the fitted mixture model will produce the posterior probability of the group memberships $Y_{jk}$, where $j = n+1,...,n+m$ and $k \in [K]$ for the unlabeled data. The categorical classification is done such that the observations belong to the class with the maximum posterior probability.  
\subsection{Model Selection}
The CEM algorithm estimates the parameters $\mu, \Sigma, \pi$. The regions or clusters centered at $\mu$ characterize the data generated from mixture of Gaussian distributions. The shapes of these regions or clusters are determined by the covariance matrices $\Sigma$. In \cite{banfield1993model}, eigen-decomposition on the covariance matrix, $\Sigma_k = \lambda_kD_kA_kD^T_k$, has geometric interpretations: $D_k$ is an orthogonal matrix consisting of the eigenvectors of $\Sigma_k$, determining the orientation of $k$-th component; $A_k$ is the diagonal matrix with the eigenvalues in the diagonal, determining the shape; and $\lambda_k$ are  the eigenvalues, determining the size of the group. For example, when all components are of the same size and spherical, the imposed constraint is $\Sigma_k = \lambda I$. If the components are spherical but of different sizes, the imposed constraint is $\Sigma_k = \lambda_k I$. 

As a result, different shapes of the mixture components lead to a different parametrization of the covariance matrix, and thus different mixture models. Here we consider the set of models listed in \cite{fraley2007model}, \cite{fraley2012mclust}, \cite{dean2006using}, \cite{russell2014up} and use the Bayesian Information criterion (BIC) for model selection.
The BIC is given by
\begin{equation}
\text{BIC}(M_i) = 2\log L_C- \ln(n+m)l,
\end{equation}
where $\log L_C$ is the maximum likelihood of the complete data, $n+m$ is the number of obeservations, and $l$ is the number of parameters. The optimal model is selected based on the maximum BIC \cite{dean2006using}, \cite{fraley2002model}, \cite{raftery1999bayes}.

\subsection{Summarizing our framework}
Our framework of model-based semi-supervised classification for dynamic Android malware analysis is summarized in Algorithm 1.



\begin{algorithm}\caption{Model-based semi-supervised classification on dynamic Android behavior}
\begin{algorithmic}
\State {\bf Goal:} Predict the label of Android APKs.
\State {\bf Input:} Android APKs. MBSS convergence stopping criterion.  
\State {\bf Step 1: Emulation.} Execute APKs, and output API names to logs.
\State {\bf Step 2: Data generation.} Process logs and extract selected API names.
\State {\bf Step 3: Feature extraction.} Engineer features from API names.
\State {\bf Step 4: MBSS} Model-based semi-supervised classification on extracted features of API logs. 
\State {\bf Step 5: Classification} Classify APKs into groups based on the maximum posterior probability from MBSS.

\end{algorithmic}
\end{algorithm}

\section{Results} \label{sec:result}
We compare MBSS with some of the most popular malware classifiers: support vector machine, k-nearest neighbor, and linear discriminant analysis. We conduct two categories of experiments: in-sample validation and out-of-sample validation. In-sample validation is the typical classification setting where a dataset is split into training and test set. A classifier is trained on the training set and expected to perform well when tested on the test set. Cross validation is usually employed to assess the performance of a classifier. In-sample classification provides the ideal classification scenario, where test data distribution is the same as the training data distribution. 

Our next set of experiments focuses on out-of-sample testing. An out-of-sample experiment uses the classifier trained and validated on the in-sample data, and predicts the labels of incoming unlabeled data. The vast majority of the unlabeled data are not guaranteed to follow the same distribution as the in-sample training data. This is a challenge for machine learning algorithms used for practical applications. We consider a classifier robust and practical when it can still achieve reasonably well performance for out-of-sample classification. All the APKs for our experiments are retrieved from VirusTotal, and the experiments are conducted in R \cite{R}, \cite{russell2014up}, \cite{fraley2012mclust}.
\subsection{Dataset}
We use the dynamic logs obtained from our Android emulator as described in Section \ref{sec:android_malware_emulator_imp}. 
The behavior data contains the filtered API calls during the execution of runs. The set of unique API calls constitute the feature space. Each data sample APK is represented by the binary feature vector denoting existence of an API call. For example, suppose the unique number of API calls is $d$, then a sample APK is represented by $x = (0,0,..,1,..,1,..,0)\in \{0,1\}^d$, where 1 denotes the existence of an API call and 0 otherwise. 



\subsection{In-sample validation} \label{subsec:in_sample_validation}
We first demonstrate that for in-sample classification, MBSS has competitive performance when compared with SVM with radial kernel, SVM with linear kernel, 3NN and LDA, all of which are most widely used for malware classification.


As in-sample dataset, we obtain 55994 Android APK dynamic logs from our emulator with 24217 benign Android APKs and 31777 malicious APKs. The in-sample malicious behaviors include stealing location, device ID, MAC information, dynamic code loading behavior, and sending the information to outside network. The label distribution is $(43\%, 57\%)$ and the chance accuracy (classification accuracy when randomly guessing) is $0.57$. We first validate the effectiveness of all five classifiers.

\begin{figure}
\center
\includegraphics[width = 8cm, height=6cm]{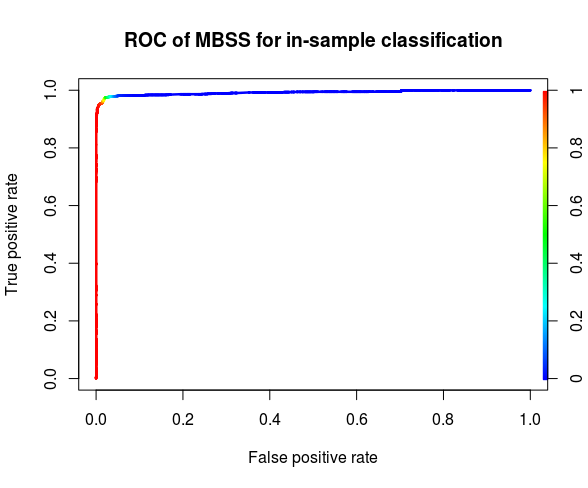}

\caption{The receiver operating curve (ROC) of MBSS for in-sample classification. The area under the curve is 0.99.}
\label{fig:in_sample_10fold_mbss}
\end{figure}

We conduct 10-fold cross validation, report the accuracy mean by averaging the accuracies and calculate the standard deviation of the accuracies across all the 10 folds validation, and report the similar metrics for false positive rates. As indicated in Table \ref{tab:all_res}, all the classifiers have competitive performance. Under this ideal classification scenario, SVM with radial and linear kernels demonstrate the best performance with accuracies at $98.8\%$ and $98.6\%$ respectively and false positive rates both at 2\%, 3NN and MBSS have similar performance of accuracy at $97.6\%$ and 97.9\% respectively and false positive rate at 3\%, while LDA shows lesser performance with accuracy at 90\% and false positive at 6\%. Figure \ref{fig:in_sample_10fold_mbss} shows the receiver operating curve (ROC) of MBSS in the 10-th fold classification and the area under the curve is 0.99.

\begin{table}
\center
\begin{tabular}{c|c|c|c|c|c|c}
\hline
Classifier & Mean ACC & Sd ACC  & Mean FP & Sd FP& DR for OOS1 & DR for OOS2 \\\hline
MBSS & 97.6\% &0.002 & 3\%  &0.004 & 90.0\% & 55.3\%\\\hline

SVM (radial)  & 98.8\% &  0.002 & 1.8\% & 0.003 &0 &0.06\%\\\hline
SVM (linear) & 98.6\% & 0.001 & 2\% & 0.003 & 90.8\% & 35.4\%\\\hline

3NN & 97.9\% & 0.001 & 3\% &0.004  & 68.4\% & NA \\ \hline
LDA & 90.0\% & 0.003 & 6\% & 0.003& 9.4\% & 33.8\%\\ \hline


\end{tabular}
\caption{Classification performance comparison for all five classifiers. All classifiers have competitive classification performance for in-sample testing. For OOS1, MBSS and SVM with linear kernel achieve the highest detection rate. For OOS2, MBSS performs significantly better than all other classifiers. Due to too many ties for 3NN, we do not report its result here.
}
\label{tab:all_res}
\end{table}




\subsection{Out-of-sample classification}

Our next experiment is out-of-sample validation. For our out-of-sample classification experiment, we apply the five classifiers on a test dataset consisting of all malicious samples, and thus our task is to detect the malicious samples. We report the detection rate (DR), which is defined by the number of correctly classified malware APKs divided by the total number of out-of-sample test data. After validating that these classifiers have high accuracy and low false positive in Section \ref{subsec:in_sample_validation}, we use them to test on incoming samples. Under this practical and realistic scenario, the test samples do not follow very similar distribution as the training samples. In this case, the classifiers with high accuracy degrade, sometimes even significantly, for out-of-sample testing. 

Recall that the in-sample malicious behaviors include retrieving phone information and sending to the network. Here we divide the out-of-sample experiments into two types. First, a strong similarity of this malicious behavior from the test set, which indicates that the distribution of the test set is similar to the distribution of a subset in the training data. Second, a weaker similarity of this malicious behavior from the test set, which indicates that the distribution of the test set is even less similar to the distribution of the training data.



\subsubsection{OOS1: Out-of-sample with malicious similarity to in-sample data}
\begin{figure}
\center
\includegraphics[width = 6cm, height=6cm]{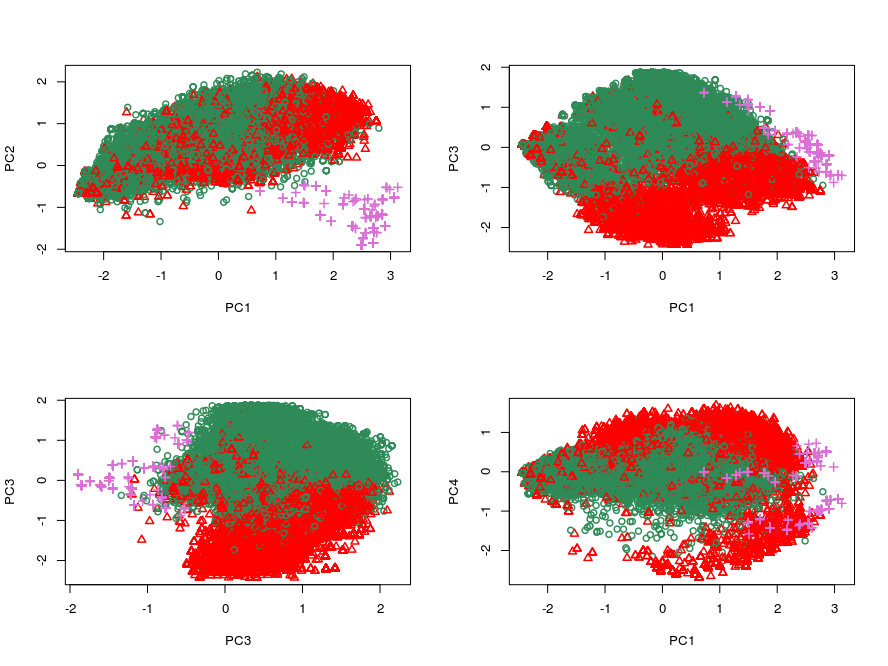}
\caption{OOS1: Scatter plot of principal components in OOS1. This out-of-sample set is similarly distributed compared to the training set. The in-sample benign files are plotted in green, the in-sample malicious files are plotted in red and the out-of-sample malicious files are plotted in magenta. (PC1, PC2)-scatter plot indicates the distributional dissimilarity between the first two principal components, as the out-of-sample data almost forms its own cluster. The rest of the PC-scatter plots show that the out-of-sample data lies in the same embedding as the in-sample data. }
\label{fig:oos1_pca_scatter}
\end{figure}

We first apply the five classifiers on a dataset of 12185 malicious APKs and report the detection rate. These out-of-sample APKs exhibit similar malicious behaviors of intercepting and sending messages without the user's consent as in the training set. To visualize this similarity, we conduct principal component analysis (PCA) and inspect the scatter plots of the first four principal components (PC). Figure \ref{fig:oos1_pca_scatter} presents the scatter plot of (PC1, PC2), (PC1, PC3), (PC2, PC3), and (PC1, PC4). The in-sample benign files are plotted in green, the in-sample malicious files are plotted in red and the out-of-sample malicious files are plotted in magenta. (PC1, PC2)-scatter plot indicates the distributional dissimilarity between the first two principal components, as the out-of-sample data almost forms its own cluster. However (PC1, PC3), (PC2, PC3), and (PC1, PC4)-scatter plots indicate that similarity still persists as the out-of-sample data falls within the clusters of the in-sample data. 

The sixth column in Table \ref{tab:all_res} demonstrates the detection rates of the five classifiers. Both SVM with linear kernel and MBSS have detection rate of 0.9, while the performance of 3NN is 0.68 and LDA has the lowest detection rate of 0.1. A dramatic performance degradation is seen for SVM with radial kernel, which went from the best in-sample performing classifier to the worst classifier with detection rate near 0. The significant degradation of LDA is due to its highly parametric nature. In this case, SVM with linear kernel and MBSS maintain relatively stable detection rate.

\begin{figure}
\center
\includegraphics[width = 6cm, height=6cm]{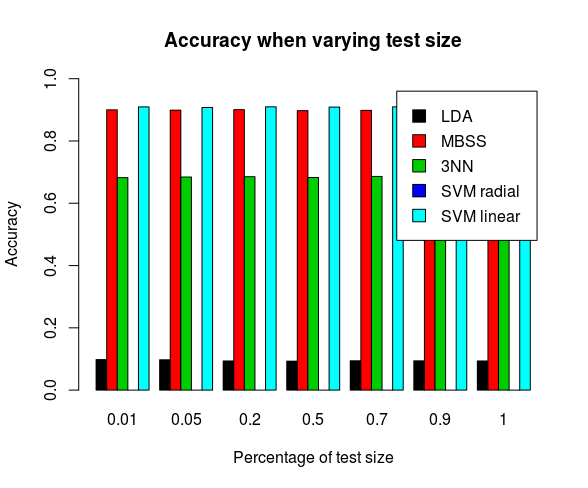}
\caption{OOS1: Detection rate as we vary the test size. MBSS and SVM linear achive the highest detection rate at 90\% while the rest of the classifiers degrade in performance.}
\label{fig:oos1_vary_train}
\end{figure}

Next, we examine the detection rate as we vary the test size. We apply the classifiers on the randomly selected $\{0.1\%, 1\%, 20\%, 50\%, 90\%\}$ of the test data with independent Monte Carlo replications at $\{50, 30, 20, 10, 5,1\}$ respectively. Monte Carlo replications are used to control the variation and thus provide better estimate of the accuracy. The detection rate is reported by averaging over the Monte Carlo replicates. As seen in Figure \ref{fig:oos1_vary_train}, both MBSS and SVM with linear kernel are superior to other classifiers. We note that the performance variation, as we increase the percentage of the test set, is negligible. This is because our data processing of binarizing the features results in a high number of replicates in the test data. Hence, the performance of the classifiers are seen relatively stable here as we vary the test size. 

\subsubsection{OOS2: Out-of-sample with dissimilar distribution to in-sample data}

Our next out-of-sample experiment applies the five classifiers onto a dataset of $11986$ malicious APKs, whose malicious behaviors primary include stealing private information, sending it to the Internet through commodity command and control (C\&C) server, but do not include dynamic code loading behavior. Compared to the training set, these APKs share similarities in the malicious behavior but have their own characteristics. Indeed, these APKs are considered a malware family that is not in the training set.

\begin{figure}
\center
\includegraphics[width = 6cm, height=6cm]{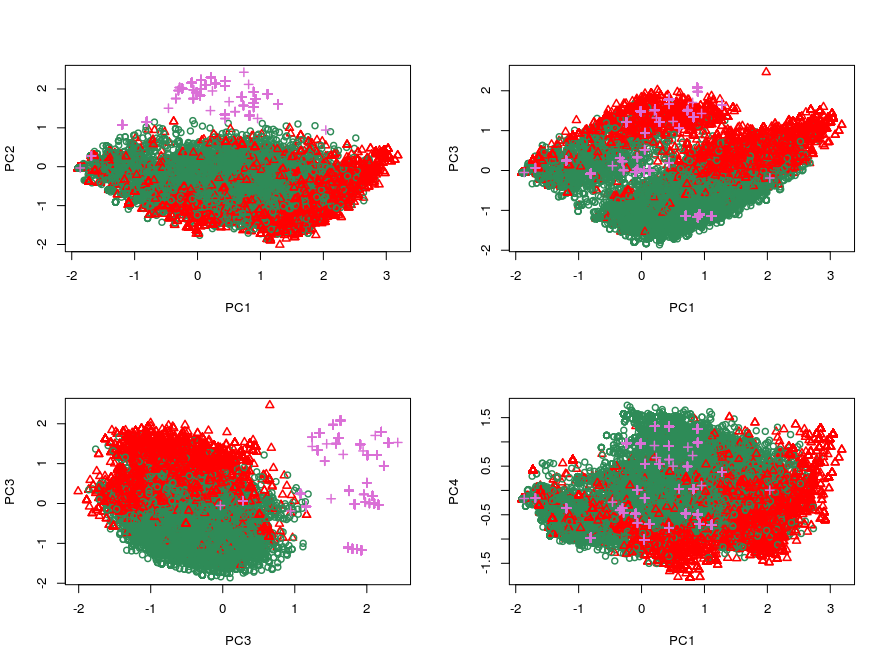}
\caption{OOS2: Scatter plot of principal components in OOS2.This out-of-sample dataset is not so similarly distributed compared to training set. The in-sample benign files are plotted in green, the in-sample malicious files are plotted in red and the out-of-sample malicious files are plotted in magenta. (PC1, PC2)-scatter plot and (PC2, PC3)-scatter plot indicate the distributional dissimilarity, as the out-of-sample data almost forms its own cluster. However (PC1, PC3) and (PC1, PC4)-scatter plots indicate there is still similarity, even though not as strong, since the out-of-sample data falls within the same embedding of the in-sample data.}
\label{fig:oos2_pca_scatter}
\end{figure}

This slight similarity in malicious behavior can been seen in the data distribution as reflected by the principal components achieved from principal component analysis. Figure \ref{fig:oos2_pca_scatter} presents the scatter plot of (PC1, PC2), (PC1, PC3), (PC2, PC3), and (PC1, PC4). The in-sample benign files are plotted in green, the in-sample malicious files are plotted in red and the out-of-sample malicious files are plotted in magenta. (PC1, PC2)-scatter plot and (PC2, PC3)-scatter plot indicate the distributional dissimilarity, as the out-of-sample data almost forms its own cluster. However (PC1, PC3) and (PC1, PC4)-scatter plots indicate there is still similarity, even though not as strong, since the out-of-sample data falls within the clusters of the in-sample data.

The last right column in Table \ref{tab:all_res} demonstrates the detection performance of the five classifiers. All classifiers degrade in performance. However SVM with linear kernel degrades in performance significantly. The result of 3NN is omitted, because it fails due to too many ties. MBSS performs significantly better than the other classifiers.

\begin{figure}
\center
\includegraphics[width = 6cm, height=6cm]{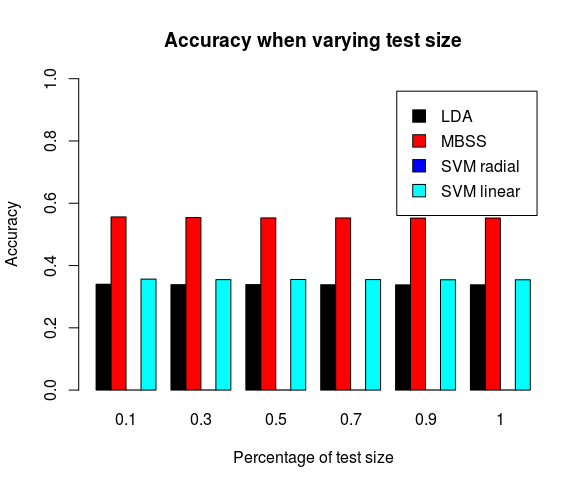}
\caption{Detection rate as we vary the test size. All classifiers degrade in performance. MBSS still performs significantly better the rest of the classifiers.}
\label{fig:oos2_vary}
\end{figure}

Next, we vary the test size to examine the out-of-sample classification performance. As we vary the test size, we apply the classifiers on the randomly selected $\{0.1\%, 1\%, 20\%, 50\%, 90\%\}$ of the test data with independent Monte Carlo replications at $\{50, 30, 20, 10, 5,1\}$ respectively. Monte Carlo replications are used to control the variation and thus provide better estimate of the accuracy. MBSS performs significantly better than the rest at all test sizes, as seen in Figure \ref{fig:oos2_vary}. The variation of the performance by these classifiers are negligible. Again this is due to our data processing, which results in many duplicates in the test data. 
 



\section{Summary and Discussion} \label{sec:conclusion}

Automated malware detection on Android platforms has focused mainly on developing signature-based methods or machine learning methods on static data, which may not capture the malicious behaviors exhibiting only during runtime. Furthermore, producing ground truth via manual labeling is costly, while a vast amount of unlabeled malware data already exists. The malware evolution process creates huge engineering problem for anti-virus researchers as they lack an efficient way to capture potential new malicious files while pruning out those clean files and known files. As a result, traditional supervised machine learning algorithms can degrade in performance.

In this paper, we demonstrate the effectiveness of using model-based semi-supervised learning (MBSS) approach on dynamic Android behavior data for Android malware detection. We show that for in-sample testing, MBSS has competitive accuracy and false positive rate compared with the most popular malware classifiers. For out-of-sample testing, MBSS produces significantly higher detection rate compared with the other classifiers in consideration. We are optimistic that the framework of semi-supervised learning for dynamic analysis is valuable and practical for anti-malware research.
investigation.

\subsection{Application on Malware Triage}
Malware triage is a known problem for security researchers. As malware evolves, derivative malware families will continuously generate new signatures that escape detection. This creates a huge engineering problem for anti-virus researchers, because they lack an efficient way to capture potential new malicious files while pruning out those clean files and known files. A more sophisticated tool to process massive amount of samples is desired by cybersecurity companies. This tool should precisely triage samples into clusters and identify a subset of highly probable malwares. Accordingly, the security experts can focus on a smaller subset and spend time more efficiently. 

One application of MBSS can be used for triaging samples and identifying a subset of highly probable malware. With limited known malwares and much more unknown samples in the field, Android malware detection models using semi-supervised machine learning approach can accurately identify subset of samples needed further investigation.

\subsection{N-gram} 
There has been an increasing interest in employing natural language processing techniques for malware detection. As a pre-processing step, $N$-gram is used to extract the features based on $N$ number of concatenated terms. With $N>1$, the feature size increases and potentially provides more discriminatory power in classification. On the other hand, we caution the usage of $N$-gram as it is not resilient against analytical attacks, which is encountered in adversarial machine learning \cite{huang2011adversarial}, a field concerning the security of machine learning. When using $(N>1)$-gram, one can break the $N$-gram feature patterns by injecting API calls, and cause the machine learning algorithm to misclassify. Here we use uni-gram, i.e., $N=1$ as a robust mitigation to address the machine learning security.
\subsection{Kernel Trick}
In this study, we compare MBSS with SVM linear kernel, SVM radial kernel, $3$NN and LDA. Essentially SVM employs kernel tricks to represent data not separable in the current dimension, and embeds it into higher dimension, such that the data becomes separable. This is the so-called kernel trick. We believe that exploring other high-dimensional representation may also help enhance the out-of-sample detection rate for SVM.
\subsection{Limitation of our framework}
Currently most detection methods for Android malware focus on developing signature-based techniques, and machine learning methods have mainly focused on static data. Static analysis provides the complete understanding of the code all at once and is relatively faster to analyze compared to dynamic analysis. However, static analysis are not resilient to obfuscation, since a significant portion of the static samples as well as the main code within the static files are usually encrypted. Simply dissembling the code statically does not expose the malicious behavior. On the other hand, using behavior data records and captures the malware behavior during dynamic execution, and thus enables better malware detection. However dynamic analysis has limited coverage, since there is no guarantee to traverse all functions of an APK at runtime. Compared to static analysis, the dynamic analysis is relatively slower.

The underlying assumption on the data distribution is mixture of Gaussians. Though mixture of Gaussians can accommodate data with different forms, and some non-Gaussian data can be approximated by a few Gaussian distributions \cite{dasgupta1998detecting}, we expect that if the data deviates from mixture of Gaussian distribution greatly, the performance may degrade. Another limitation is on high-dimensional data. The number of parameters grow as the square of the feature dimension. In this case, dimension reduction methods such as principal component analysis is desired.



\bibliographystyle{plain}
\bibliography{android.bib}

\begin{thebibliography}{10}

\bibitem{adb}
Android debug bridge.
\newblock \url{https://developer.android.com/studio/command-line/adb.html}.

\bibitem{droidbox}
Droidbox.
\newblock https://github.com/pjlantz/droidbox.

\bibitem{droidmon}
Droidmon.
\newblock https://github.com/idanr1986/droidmon.

\bibitem{gartner}
Gartner says five of top 10 worldwide mobile phone vendors increased sales in
  second quarter of 2016.
\newblock \url{http://www.gartner.com/newsroom/id/3415117}.
\newblock Egham, UK, August 19, 2016.

\bibitem{Google}
Google play served 65 billion downloads in 2015 alone.
\newblock
  \url{https://www.androidheadlines.com/2016/05/google-play-served-65-billion-downloads-2015-alone.html}.
\newblock May 18, 2016.

\bibitem{eagleeye}
Mindmac/androideagleeye.
\newblock \url{https://github.com/MindMac/AndroidEagleEye}.

\bibitem{mc}
Mobile threat report.
\newblock
  \url{https://www.mcafee.com/us/resources/reports/rp-mobile-threat-report-2016.pdf}.

\bibitem{nokia}
Nokia malware report shows surge in mobile device infections in 2016.
\newblock
  \url{http://www.nokia.com/en_int/news/releases/2016/09/01/nokia-malware-report-shows-surge-in-mobile-device-infections-in-2016}.

\bibitem{AppBrain}
Number of android applications.
\newblock
  \url{http://web.archive.org/web/20170210051327/https:/www.appbrain.com/stats/number-of-android-apps}.
\newblock February 9, 2017.

\bibitem{exposed}
rovo89/xposed.
\newblock \url{https://github.com/rovo89/xposed}.

\bibitem{idc}
Smartphone os market share, 2016 q3.
\newblock \url{http://www.idc.com/promo/smartphone-market-share/os}.

\bibitem{aafer2013droidapiminer}
Yousra Aafer, Wenliang Du, and Heng Yin.
\newblock Droidapiminer: Mining api-level features for robust malware detection
  in android.
\newblock In {\em International Conference on Security and Privacy in
  Communication Systems}, pages 86--103. Springer, 2013.

\bibitem{afonso2016going}
Vitor Afonso, Antonio Bianchi, Yanick Fratantonio, Adam Doup{\'e}, Mario
  Polino, Paulo de~Geus, Christopher Kruegel, and Giovanni Vigna.
\newblock Going native: Using a large-scale analysis of android apps to create
  a practical native-code sandboxing policy.
\newblock In {\em Proceedings of the Annual Symposium on Network and
  Distributed System Security (NDSS)}, 2016.

\bibitem{banfield1993model}
Jeffrey~D Banfield and Adrian~E Raftery.
\newblock Model-based gaussian and non-gaussian clustering.
\newblock {\em Biometrics}, pages 803--821, 1993.

\bibitem{bensmail1996regularized}
Halima Bensmail and Gilles Celeux.
\newblock Regularized gaussian discriminant analysis through eigenvalue
  decomposition.
\newblock {\em Journal of the American statistical Association},
  91(436):1743--1748, 1996.

\bibitem{celeux1995gaussian}
Gilles Celeux and G{\'e}rard Govaert.
\newblock Gaussian parsimonious clustering models.
\newblock {\em Pattern recognition}, 28(5):781--793, 1995.

\bibitem{chapelle2005semi}
Olivier Chapelle and Alexander Zien.
\newblock Semi-supervised classification by low density separation.
\newblock In {\em AISTATS}, pages 57--64, 2005.

\bibitem{chen2014stochastic}
Li~Chen and Matthew Patton.
\newblock Stochastic blockmodeling for online advertising.
\newblock {\em Proceedings of the Twenty-Ninth AAAI Conference on Artificial
  Intelligence}, 2015.

\bibitem{chen2016robust}
Li~Chen, Cencheng Shen, Joshua~T Vogelstein, and Carey~E Priebe.
\newblock Robust vertex classification.
\newblock {\em IEEE transactions on pattern analysis and machine intelligence},
  38(3):578--590, 2016.

\bibitem{cortes1995support}
Corinna Cortes and Vladimir Vapnik.
\newblock Support-vector networks.
\newblock {\em Machine learning}, 20(3):273--297, 1995.

\bibitem{cover1967nearest}
Thomas Cover and Peter Hart.
\newblock Nearest neighbor pattern classification.
\newblock {\em IEEE transactions on information theory}, 13(1):21--27, 1967.

\bibitem{dasgupta1998detecting}
Abhijit Dasgupta and Adrian~E Raftery.
\newblock Detecting features in spatial point processes with clutter via
  model-based clustering.
\newblock {\em Journal of the American Statistical Association},
  93(441):294--302, 1998.

\bibitem{dean2006using}
Nema Dean, Thomas~Brendan Murphy, and Gerard Downey.
\newblock Using unlabelled data to update classification rules with
  applications in food authenticity studies.
\newblock {\em Journal of the Royal Statistical Society: Series C (Applied
  Statistics)}, 55(1):1--14, 2006.

\bibitem{dempster1977maximum}
Arthur~P Dempster, Nan~M Laird, and Donald~B Rubin.
\newblock Maximum likelihood from incomplete data via the em algorithm.
\newblock {\em Journal of the royal statistical society. Series B
  (methodological)}, pages 1--38, 1977.

\bibitem{devroye2013probabilistic}
Luc Devroye, L{\'a}szl{\'o} Gy{\"o}rfi, and G{\'a}bor Lugosi.
\newblock {\em A probabilistic theory of pattern recognition}, volume~31.
\newblock Springer Science \& Business Media, 2013.

\bibitem{fraley1998many}
Chris Fraley and Adrian~E Raftery.
\newblock How many clusters? which clustering method? answers via model-based
  cluster analysis.
\newblock {\em The computer journal}, 41(8):578--588, 1998.

\bibitem{fraley2002model}
Chris Fraley and Adrian~E Raftery.
\newblock Model-based clustering, discriminant analysis, and density
  estimation.
\newblock {\em Journal of the American statistical Association},
  97(458):611--631, 2002.

\bibitem{fraley2007model}
Chris Fraley, Adrian~E Raftery, et~al.
\newblock Model-based methods of classification: using the mclust software in
  chemometrics.
\newblock {\em Journal of Statistical Software}, 18(6):1--13, 2007.

\bibitem{fraley2012mclust}
Chris Fraley, Adrian~E. Raftery, Thomas~Brendan Murphy, and Luca Scrucca.
\newblock {\em mclust Version 4 for R: Normal Mixture Modeling for Model-Based
  Clustering, Classification, and Density Estimation}, 2012.

\bibitem{huang2011adversarial}
Ling Huang, Anthony~D Joseph, Blaine Nelson, Benjamin~IP Rubinstein, and
  JD~Tygar.
\newblock Adversarial machine learning.
\newblock In {\em Proceedings of the 4th ACM workshop on Security and
  artificial intelligence}, pages 43--58. ACM, 2011.

\bibitem{jebara1998maximum}
Tony Jebara and Alex Pentland.
\newblock Maximum conditional likelihood via bound maximization and the cem
  algorithm.
\newblock In {\em Proceedings of the 11th International Conference on Neural
  Information Processing Systems}, pages 494--500. MIT Press, 1998.

\bibitem{johnson1967hierarchical}
Stephen~C Johnson.
\newblock Hierarchical clustering schemes.
\newblock {\em Psychometrika}, 32(3):241--254, 1967.

\bibitem{lu2012chex}
Long Lu, Zhichun Li, Zhenyu Wu, Wenke Lee, and Guofei Jiang.
\newblock Chex: statically vetting android apps for component hijacking
  vulnerabilities.
\newblock In {\em Proceedings of the 2012 ACM conference on Computer and
  communications security (CCS)}, pages 229--240. ACM, 2012.

\bibitem{lyzinski2015spectral}
Vince Lyzinski, Daniel~L Sussman, Donniell~E Fishkind, Henry Pao, Li~Chen,
  Joshua~T Vogelstein, Youngser Park, and Carey~E Priebe.
\newblock Spectral clustering for divide-and-conquer graph matching.
\newblock {\em Parallel Computing}, 47:70--87, 2015.

\bibitem{mariconti2016mamadroid}
Enrico Mariconti, Lucky Onwuzurike, Panagiotis Andriotis, Emiliano
  De~Cristofaro, Gordon Ross, and Gianluca Stringhini.
\newblock Mamadroid: Detecting android malware by building markov chains of
  behavioral models.
\newblock 2016.

\bibitem{mclachlan2004finite}
Geoffrey McLachlan and David Peel.
\newblock {\em Finite mixture models}.
\newblock John Wiley \& Sons, 2004.

\bibitem{moon1996expectation}
Todd~K Moon.
\newblock The expectation-maximization algorithm.
\newblock {\em IEEE Signal processing magazine}, 13(6):47--60, 1996.

\bibitem{ng2001spectral}
Andrew~Y Ng, Michael~I Jordan, Yair Weiss, et~al.
\newblock On spectral clustering: Analysis and an algorithm.
\newblock In {\em NIPS}, volume~14, pages 849--856, 2001.

\bibitem{R}
{R Core Team}.
\newblock {\em R: A Language and Environment for Statistical Computing}.
\newblock R Foundation for Statistical Computing, Vienna, Austria, 2016.

\bibitem{raftery1999bayes}
Adrian~E Raftery.
\newblock Bayes factors and bic: Comment on “a critique of the bayesian
  information criterion for model selection”.
\newblock {\em Sociological Methods \& Research}, 27(3):411--427, 1999.

\bibitem{russell2014up}
Niamh Russell, Laura Cribbin, and Thomas~Brendan Murphy.
\newblock {\em upclass: Updated Classification Methods using Unlabeled Data},
  2014.
\newblock R package version 2.0.

\bibitem{safavian1991survey}
S~Rasoul Safavian and David Landgrebe.
\newblock A survey of decision tree classifier methodology.
\newblock {\em IEEE transactions on systems, man, and cybernetics},
  21(3):660--674, 1991.

\bibitem{seo2016flexdroid}
Jaebaek Seo, Daehyeok Kim, Donghyun Cho, Taesoo Kim, and Insik Shin.
\newblock Flexdroid: Enforcing in-app privilege separation in android.
\newblock In {\em Proceedings of the 2016 Annual Network and Distributed System
  Security Symposium (NDSS)}, pages 1--53, 2016.

\bibitem{shabtai2010google}
Asaf Shabtai, Yuval Fledel, Uri Kanonov, Yuval Elovici, Shlomi Dolev, and
  Chanan Glezer.
\newblock Google android: A comprehensive security assessment.
\newblock {\em IEEE Security \& Privacy}, 8(2):35--44, 2010.

\bibitem{wong2016intellidroid}
Michelle~Y Wong and David Lie.
\newblock Intellidroid: A targeted input generator for the dynamic analysis of
  android malware.
\newblock In {\em Proceedings of the Annual Symposium on Network and
  Distributed System Security (NDSS)}, 2016.

\bibitem{wright2009robust}
John Wright, Allen~Y Yang, Arvind Ganesh, S~Shankar Sastry, and Yi~Ma.
\newblock Robust face recognition via sparse representation.
\newblock {\em IEEE transactions on pattern analysis and machine intelligence},
  31(2):210--227, 2009.

\bibitem{xue2017auditing}
Yinxing Xue, Guozhu Meng, Yang Liu, Tian~Huat Tan, Hongxu Chen, Jun Sun, and
  Jie Zhang.
\newblock Auditing anti-malware tools by evolving android malware and dynamic
  loading technique.
\newblock {\em IEEE Transactions on Information Forensics and Security}, 2017.

\end{thebibliography}
\end{document}